\documentclass[%
 reprint,
superscriptaddress,
 amsmath,amssymb,
 prx,
]{revtex4-2}

\usepackage{graphicx}
\usepackage{dcolumn}
\usepackage{bm}
\usepackage{hyperref}
\usepackage{placeins}
\usepackage{physics}
\usepackage{babel}
\usepackage{xcolor}
\usepackage{dsfont}
\usepackage[caption=false]{subfig}
\usepackage{amsthm}
\usepackage{braket}
\usepackage{optidef}
\usepackage{algorithm2e}
\usepackage[toc,page]{appendix}
\usepackage{hhline}
\usepackage{tikz}
\usetikzlibrary{quantikz2}
\RestyleAlgo{ruled}

\makeatletter
\let\ORIbbl@fixname\bbl@fixname
\def\bbl@fixname#1{%
  \@ifundefined{languagealias@\expandafter\string#1}
    {\ORIbbl@fixname#1}
    {\edef\languagename{\@nameuse{languagealias@#1}}}%
}
\newcommand{\definelanguagealias}[2]{%
  \@namedef{languagealias@#1}{#2}%
}
\makeatother

\definelanguagealias{en}{english}
\definelanguagealias{EN}{english}
\definelanguagealias{en-GB}{english}

\newcommand{\btheta}{{\bm{\theta}}}
\newcommand{\bx}{{\bm{x}}}

\newcommand{\bphi}{\bm{\phi}}

\newcommand{\U}[1]{\mathrm{U}(#1)}
\newcommand{\SU}[1]{\mathrm{SU}(#1)}
\newcommand{\su}[1]{\mathfrak{su}(#1)}

\begin{document}

\preprint{APS/123-QED}

\title{Geodesic Algorithm for Unitary Gate Design with Time-Independent Hamiltonians}

\author{Dylan Lewis}
\affiliation{Department of Physics and Astronomy, University College London, London WC1E 6BT, United Kingdom}

\author{Roeland Wiersema}
\affiliation{Vector Institute, MaRS  Centre,  Toronto,  Ontario,  M5G  1M1,  Canada}
\affiliation{Department of Physics and Astronomy, University of Waterloo, Ontario, N2L 3G1, Canada}

\author{Juan Carrasquilla}
\affiliation{Institute for Theoretical Physics, ETH Zürich, 8093, Switzerland}
\affiliation{Vector Institute, MaRS  Centre,  Toronto,  Ontario,  M5G  1M1,  Canada}

\author{Sougato Bose}
\affiliation{Department of Physics and Astronomy, University College London, London WC1E 6BT, United Kingdom}


\begin{abstract}
Larger multi-qubit quantum gates allow shallower, more efficient quantum circuits, which could decrease the prohibitive effect of noise on algorithms for noisy intermediate-scale quantum (NISQ) devices and fault-tolerant error correction schemes.
Such multi-qubit gates can potentially be generated by time-independent Hamiltonians comprising only physical (one- and two-local) interaction terms. Here, we present an algorithm that finds the time-independent Hamiltonian for a target quantum gate on $n$ qubits by using the geodesic on the Riemannian manifold of $\mathrm{SU}(2^n)$. Differential programming is used to determine how the Hamiltonian should be updated in order to follow the geodesic to the target unitary as closely as possible. 
We show that our geodesic algorithm outperforms gradient descent methods for standard multi-qubit gates such as Toffoli and Fredkin. The geodesic algorithm is then used to find previously unavailable multi-qubit gates implementing high fidelity parity checks, which could be used in a wide array of quantum codes and increase the clock speed of fault-tolerant quantum computers. The geodesic algorithm is demonstrated on an example relevant to current experimental hardware, illustrating a circuit speed up.

\end{abstract}

\maketitle

\section{Introduction}
Quantum computing’s potential to solve problems that are currently intractable for classical computers relies on the availability and accurate application of quantum logic gates.  These  quantum operations remain the principal component of the circuit model for quantum computing  and have been extensively studied within various quantum computing hardware platforms and algorithms~\cite{barenco_elementary_1995, nielsen_quantum_2010,porras_effective_2004,kim_quantum_2011,blatt_quantum_2012,wall_boson-mediated_2017,kiely_relationship_2017,morgado_quantum_2021,lewis_ion_2023}. Current methods to implement multi-qubit quantum gates involve performing them as a sequence of one and two qubit gates, which require switching different Hamiltonians on for specific durations of time: this is equivalent to the time-dependent control of Hamiltonians to engineer a specific unitary. The control pulses required to change the Hamiltonian during this process are one of the many sources of noise that decrease the gate fidelity, which limits the usefulness of currently available quantum hardware. Ion trap quantum simulators, used for modelling condensed matter systems, and problems in high-energy physics, have recently seen advances in their capability to engineer a range of Hamiltonian terms~\cite{monroe_programmable_2021,foss-feig_progress_2024}. 
Quantum simulators generally implement time-independent Hamiltonians. A useful quantum simulator has the capability to tune the interactions and local fields. Numerous quantum spin models have been implemented on a variety of hardware platforms, such as ion traps~\cite{foss-feig_progress_2024} and Rydberg atoms~\cite{bernien_probing_2017, ebadi_quantum_2021}. Anisotropic Heisenberg models, with varying degrees of interaction tuneability, have been realised in Rydberg atoms~\cite{pinheiro_xyz_2013, jepsen_transverse_2021, kim_realization_2024}, and in ion traps with Floquet engineering~\cite{kranzl_observation_2023}. The results here address the question of how these engineered quantum simulators could also be used for quantum computation.

The power of quantum algorithms for noisy intermediate-scale quantum (NISQ) devices~\cite{Preskill2018, chen_complexity_2022} is ultimately limited by the accumulation of noise. Candidates for quantum advantage in NISQ devices, such as variational quantum algorithms (VQAs)~\cite{cerezo_variational_2021}, may require excessively deep circuits if only a limited set of entangling gates are experimentally available~\cite{linke_experimental_2017}.
Going beyond NISQ computing, fault-tolerant quantum computing will benefit from more accurate, complex gates. In particular, in quantum error correction, the unitary evolution of a parity check could be condensed to only a single quantum gate, which would greatly reduce the optimal clock speed of the quantum processor and lead to more accurate syndrome checks. For example, the Bacon-Shor and Floquet code~\cite{li_direct_2018,hastings_dynamically_2021} both require weight two parity-checks, and there are subsystem codes that require weight three parity-checks~\cite{bravyi_subsystem_2013}. Additionally, the toric code and surface code require weight four parity checks~\cite{kitaev_quantum_1997, bravyi_quantum_1998, kitaev_fault-tolerant_2003}. Finally, there are more complex codes that require higher-weight parity checks, e.g. color codes~\cite{chamberland_triangular_2020} and weight six parity checks~\cite{srivastava_xyz2_2022} in the honeycomb XYZ$^2$ code. 

Quantum optimal control methods can be used to design time-dependent Hamiltonian terms to implement quantum gates~\cite{werschnik_quantum_2007, glaser_training_2015, khaneja_optimal_2005, doria_optimal_2011, caneva_chopped_2011,sauvage_optimal_2022}. However, the external controllers themselves can introduce noise~\cite{muller_information_2022,aroch_mitigating_2024}. The sources of noise for time-dependent control depend on the hardware platform. In superconducting qubits and quantum dots, since the system's energy levels are swept through an avoided crossing, time dependence in the Hamiltonian introduces unintended transition probabilities, as described by the Landau-Zener model~\cite{nyisomeh_landauzener_2020, mcewen_removing_2021}. In piecewise Hamiltonians, control fields have to be changed rapidly. A timing error occurs for each rapid change in the Hamiltonian~\cite{ellert-beck_power-optimized_2024}, which means that multiple piecewise Hamiltonians accumulate more error than a single time-independent Hamiltonian. The ability to perform sudden changes in the Hamiltonian is limited because of the bandwidth of the acousto-optic modulators, or electro-optic modulators, used to shift the laser frequencies. Fewer piecewise Hamiltonians, and using time-independent Hamiltonians wherever possible, would minimise error from a limited bandwidth. Even if these errors are small, there is a calibration advantage in using time-independent Hamiltonians. Real quantum systems may deviate slightly from the effective Hamiltonian models used for quantum gate design~\cite{islam_emergence_2013}. Unlike optimal control pulses, time-independent Hamiltonians are more straightforward to calibrate since all the model error comes from adjusting Hamiltonian strengths, as in quantum simulation~\cite{lewis_ion_2023}, rather than the complicated case of control functions for time-dependent Hamiltonians. Large multi-qubit gates with time-independent Hamiltonians could therefore lead to increased gate fidelities. The approach of minimising control could also substantially reduce gate execution time by utilising the full Hamiltonian flexibility offered by quantum simulators. This would further reduce noise, as the system has less time during gate operation to interact with the environment.
Although there are some examples of quantum gates with time-independent Hamiltonians~\cite{benjamin_quantum_2003,Banchi2016,Eloie2018,Innocenti2018,Innocenti2020,majumder_variational_2022}, an efficient method specifically designed for this problem has not yet been found. 

Finding the Hamiltonian that generates a specific gate under hardware constraints is numerically challenging. 
We offer a solution to the problem of generating complex multi-qubit gates from time-independent Hamiltonians through the lens of differential geometry of the Lie group structure of quantum gates. Some geometric techniques have previously been crucial for understanding quantum circuit complexity~\cite{nielsen_geometric_2005, nielsen_quantum_2006, bhattacharyya_renormalized_2020}.
In this work, we develop an algorithm that finds the Hamiltonian’s couplings using the geodesics on the Riemannian manifold of the $\SU{N}$ group and demonstrate its efficiency by comparison to gradient descent techniques for the generation of Toffoli and Fredkin gates. Furthermore, we use the algorithm to generate weight-$k$ parity checks necessary for a wide array of quantum error correcting codes. We find that our geodesic algorithm is significantly more efficient than stochastic gradient descent and gradient descent algorithms for finding a restricted generating Hamiltonian of a desired unitary gate.   

\section{\label{sec:preliminaries}Preliminaries}
A quantum gate $U$ with $n$ qubits is an element of the unitary group $\U{N}$, such that $U U^\dagger = U^\dagger U = I$, where $N = 2^n$. If we ignore the global phase, $U$ is an element of the special unitary group $\SU{N}$. The group $\SU{N} \subset \U{N}$ consists of all $N \times N$ unitary matrices with determinant 1 and has dimension $N^2-1$. Since $\SU{N}$ is a Lie group it is also a smooth manifold. For simply connected Lie groups, such as $\SU{N}$, a Lie group element can be generated by a Lie algebra element via the Lie group-Lie algebra correspondence. This correspondence is given by the exponential map $g = e^{A}$, which connects Lie group element $g \in \SU{N}$ to the Lie algebra element $A \in \su{N}$. The Lie algebra $\su{N} = \set{\Omega \in \mathbb{C}^{N\times N} | \Omega = -\Omega^\dagger, \Tr{\Omega} = 0}$ is a vector space of dimension $N^2 - 1$ that consists of traceless skew-Hermitian matrices and is closed under the commutator $[A, B] = AB - BA \in \su{N}$.  

\begin{figure}
    \centering
    \includegraphics[scale=1.3]{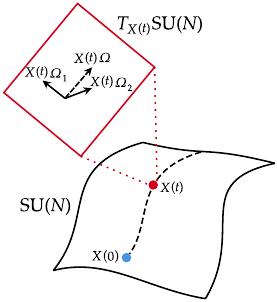}
    \caption{A curve in $\SU{N}$ is parameterised by a continuous variable $t\in\mathbb{R}$. At the point $X(t)$ there is a tangent space of directional derivatives $T_{X(t)}\SU{N}$ with direction $\Omega = \Omega(t)$.}
    \label{fig:tangent_space_diagram}
\end{figure}

We can consider the set of all Pauli words of length $n$, excluding the word of identity matrices, as a basis for traceless Hermitian matrices,
\begin{align}
    \mathcal{P} = \{\sigma^{\alpha_1} \otimes \sigma^{\alpha_2} \otimes ... \otimes \sigma^{\alpha_n} \} \setminus \{\sigma^0 \otimes \sigma^0 \otimes ... \otimes \sigma^0 \},
\end{align}
where $\alpha_i = \{0, 1, 2, 3\}$, corresponding to $\sigma^{\alpha_i} = \{I, X, Y, Z\}$. A basis of the skew-Hermitian matrices of $\su{N}$ is then given by multiplying a Pauli word with the imaginary unit $i$. For a fixed time $t=1$ (which can also be a rescaling of Hamiltonian strengths) and ignoring any global fields, Hamiltonians can then be identified as the vectors in the Lie algebra 
\begin{align}
    H(\btheta) = \sum_{j=1}^{N^2 -1} \theta_j G_j = \btheta \cdot \boldsymbol{G},
\end{align}
where $\btheta = (\theta_1, \theta_2, ... , \theta_{N^2-1}) \in \mathbb{R}^{N^2-1}$, and $G_j$ is the $j$th element of the set of Pauli words $\mathcal{P}$ ordered lexicographically. The parameter vector $\btheta$ gives the coordinates in the Lie algebra that generates the unitary $U(\btheta) \in \SU{N}$ via the exponential map:
\begin{align}
    U(\btheta) = \exp{i H(\btheta)}\label{eq:U_theta}
\end{align}
The tangent space of a smooth manifold at a specific point is a vector space formed by the set of all directional derivatives through that point. The directional derivative of a curve $X(t)$ on $\SU{N}$ that passes through the identity element at $t=0$ is
\begin{align}
    \dot{X}(0) := \frac{dX(t)}{dt}\bigg\vert_{t=0} = \Omega,
\end{align}
where $\Omega \in \su{N}$.  
Hence the tangent space of the Lie group at the identity element is given by the Lie algebra. We can move from the tangent space at the identity to the tangent space at any point $U$ by left multiplication with $U$,
\begin{align}
    T_U\SU{N} = \set{U\Omega | \Omega \in \su{N}},
\end{align}
see Fig.~\ref{fig:tangent_space_diagram}. For the parametrisation of $U$ given in Eq~\eqref{eq:U_theta}, these directional derivatives are thus given by,
\begin{align}
    \label{eq:partial_U_effective_generator}
    \frac{\partial U(\btheta)}{\partial \theta_l} = U(\btheta) \Omega_l(\btheta),
\end{align}
where there is an \emph{effective generator}, $\Omega_l$, of the directional derivative in the tangent space for each Lie algebra component $\theta_l$. 
The directional derivatives $\Omega_l$ can be calculated classically by using a differentiable programming framework such as JAX~\cite{jax2018github}, which implements a differentiable version of the exponential map~\cite{wiersema_here_2023}. 

We can equip the smooth manifold of $\SU{N}$ with the metric $g(x,y) = \Tr{x^\dagger y}/N$ to turn it into a Riemannian manifold. The metric gives a notion of distance between elements of the $\SU{N}$ group. 

A \emph{geodesic} on $\SU{N}$ is a one-parameter subgroup of the Lie group, $X(t)$, parameterised by $t$ that represents the shortest path between two unitaries. The geodesic between two unitaries $U(\btheta)$ and $V$ can be constructed as follows. Let $X(t)$ be a one-parameter subgroup in $\SU{N}$ such that $U(\btheta)X(0) = U(\btheta)$ and $U(\btheta)X(1) = V$. The geodesic between $U(\btheta)$ and $V$ can then be straightforwardly constructed as
\begin{align}
    X(t) = e^{i\Gamma t},
\end{align}
where $\Gamma \in T_{X(t)}\SU{N}$ is the direction of the geodesic on the tangent space given by
\begin{align}
    \Gamma = -i \log(U(\btheta)^\dagger V).
\end{align}

We now introduce the concept of a limited set of Hamiltonian terms -- a restriction of the Lie algebra. The basis Pauli words of the Lie algebra, $G_l$, are each possible terms of a Hamiltonian. However, Hamiltonians that are physically realisable may not include all possible terms. For example, only one- and two-body interactions may be accessible in an experimental setup. The set of these terms we define as $\mathcal{H} \subset \mathcal{P}$. We can define a binary function over a restricted set,
\begin{align}
    \bar{R}_{\mathcal{H}}(G_l) = \begin{cases}
        1, & \text{if $G_l$ is in $\mathcal{H}$}, \\
        0 & \text{otherwise}.
    \end{cases}
\end{align}
This function can be represented with a diagonal matrix that acts on Lie algebra vectors $\btheta$ and has elements
\begin{align}
    (R_{\mathcal{H}})_{ij} = \delta_{ij} \bar{R}_{\mathcal{H}}(G_i).
\end{align}
We use the notation $\btheta$ to refer to unrestricted vectors, and $\bphi$ to refer to restricted vectors where many components of the vector may have been forced to 0,
\begin{align}
    \bphi = R_{\mathcal{H}} \btheta. \label{eq:restriction}
\end{align}
We will produce multiple sets of parameters via an optimisation algorithm. The initial parameter vector is denoted by $\bphi^{(0)}$ and subsequent parameter vectors are $\bphi^{(m)}$ with $0<m\leq M$. 
The geodesic of the $m$th parameter vector $\bphi^{(m)}$ is $X^{(m)}(t)$ and its generator is $\Gamma^{(m)}$. The norm of the parameter difference between optimisation steps is the step size, 
\begin{align}
    \delta^{(m)} = \Vert\bm{\delta\theta}^{(m)}\Vert =  \Vert\bm{\theta}^{(m+1)} - \bm{\theta}^{(m)}\Vert,
\end{align}
where $\Vert \cdot \Vert$ is the Euclidean 2-norm.

Finally, we include a constraint on the Hamiltonians, as discussed in Ref.~\cite{Innocenti2020}, that can generate a specific $U$ at a fixed time $t=1$. All Hamiltonians that generate $U$ must commute. Given a Hamiltonian of a restricted parameter vector $H(\bphi)$ that generates the quantum gate $U$, we have the necessary condition
\begin{align}
    \label{eq:commutator_condition}
    [H(\bphi), \log(U)] = 0. 
\end{align}
This commutator condition further constrains the space of possible Hamiltonians within the restriction $\mathcal{H}$.

\section{\label{sec:algorithm}Algorithm}
Consider a target unitary $V$ and a parametrised unitary $U(\bphi) = \exp{iH(\bphi)}$, where $\bphi$ is a restricted vector due to the restriction set $\mathcal{H}\subset \mathcal{P}$ of the generating Hamiltonian $H(\bphi)$. We can use the metric $g$ to define the \emph{unitary infidelity}
\begin{align}
    \mathcal{I}(\bphi) = 1 - \frac{1}{N}\Tr{U^\dag(\bphi) V} \label{eq:infid},
\end{align}
which satisfies $\mathcal{I}(\bphi)\in [0,1]$. We aim to minimise $\mathcal{I}(\bphi)$ and find a set of parameters $\bphi$ that produces a unitary $U(\bphi)$ that is as close as possible to the target $V$. A naive approach that calculates the gradient of Eq.~\eqref{eq:infid} consequently updates  the parameters (see App. \ref{app:jaxgd}) will quickly get stuck in a local minimum and provide a sub-optimal approximation of $V$. Our proposed algorithm adapts this scheme by utilizing geodesic information and gradients on the group manifold to rapidly converge to an accurate solution.

The idea is to use the effective generators $\Omega_l(\bphi^{(m)})$ and the generator of the geodesic $\Gamma^{(m)}$ at $U(\bphi^{(m)})$ to determine the best way to update the parameters $\bphi^{(m+1)} = \bphi^{(m)} + \bm{\delta\phi}^{(m)}$. At each optimisation step, we want to update the parameters such that the resulting unitary $U(\bphi^{(m+1)})$ is closer to the target unitary $V$. This can be achieved by updating the parameters such that they follow (as closely as possible) the geodesic $X^{(m)}(t)$ towards the target $V$. Consider a small step along the geodesic by $\delta t$,
\begin{align}
    U(\btheta^{(m+1)}) &= U(\bphi^{(m)}) X(\delta t) \\
    &= U(\bphi^{(m)}) e^{i \Gamma^{(m)} \delta t } \\
    &= U(\bphi^{(m)}) + i  U(\bphi^{(m)}) \Gamma^{(m)} \delta t + O(\delta t^2). \label{eq:next_U_geodesic}
\end{align}
Notice how the geodesic can take us from a unitary given by a restricted parameter vector $\bphi^{(m)}$ to a unitary with a parameter vector that has components in any direction $\btheta^{(m+1)}$. 

\begin{figure*}[t]
    \centering
    \includegraphics[scale=1.]{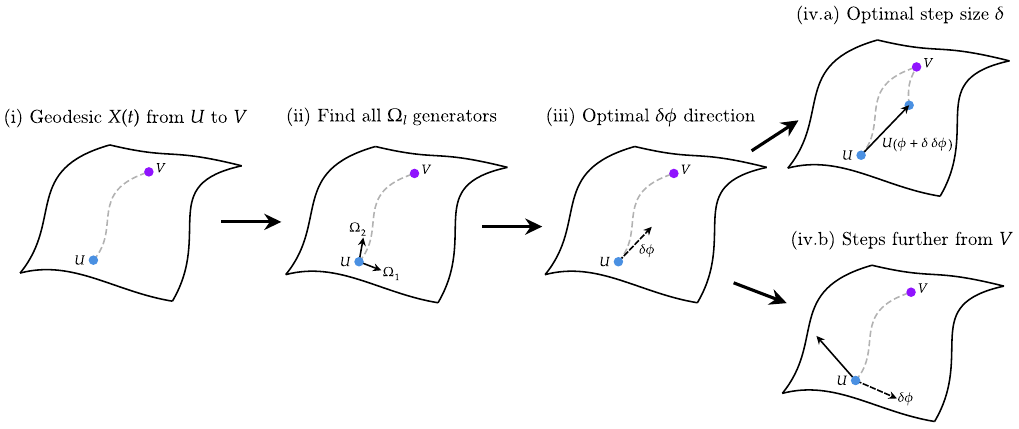}
    \caption{Illustration of the update procedure at step $m$ on the Riemannian manifold of the $\SU{N}$ group: (i) Find the geodesic from current unitary $U(\bphi^{(m)})$ to the target $V$; (ii) Find all effective generators $\Omega_j(\bphi^{(m)})$; (iii) Solve the convex optimisation problem to give the update step $\bm{\delta\phi}^{(m)}$; (iv.a) Use line search to optimise the update step size $\delta = \Vert \bm{\delta\phi}^{(m)}\Vert$; (iv.b) If the solution is currently a local minimum, use the Gram-Schmidt procedure to step away from the local minimum.}
    \label{fig:strategy}
\end{figure*}
A small change in parameter vectors, $\vert \bm{\delta \phi}\vert = \delta$, can be expanded as
\begin{align}
    U(\bphi^{(m+1)}) &= U(\bphi^{(m)}+\bm{\delta\phi}^{(m)}) \\
    &= U(\bphi^{(m)}) +\delta\phi^{(m)}_j \frac{\partial U(\bphi^{(m)})}{\partial \phi_j}\bigg\vert_{\bphi^{(m)}} + O(\delta^2),
\end{align}
with $\delta\phi_j^{(m)}$ the $j$th element of the vector $\bm{\delta\phi}^{(m)} = \bphi^{(m+1)} - \bphi^{(m)}$ and repeated indices are summed over.
Up to first order, we find
\begin{equation}
    U(\bphi^{(m+1)}) \approx U(\bphi^{(m)}) +  U(\bphi^{(m)}) \delta\phi^{(m)}_j \Omega_j(\bphi^{(m)}) , \label{eq:next_U_omega}
\end{equation}
where we have used Eq.~\eqref{eq:partial_U_effective_generator}.
We want to find the components of $\bm{\delta \phi}^{(m)}$ such that $U(\bphi^{(m+1)})$ is closer to the target $V$. We can compute all $\Omega_j(\bphi^{(m)})$ using differentiable programming and the geodesic with $\Gamma^{(m)} = - i \log(U(\bphi^{(m)})^\dagger V)$. By comparison of the first order terms in Eqs.~\eqref{eq:next_U_geodesic} and~\eqref{eq:next_U_omega}, we require 
\begin{align}
    \delta\phi^{(m)}_j \Omega_j(\bphi^{(m)}) = i \Gamma^{(m)}, \label{eq:Omega_geodesic_equivalence}
\end{align}
where repeated index $j$ is summed over and $\delta t$ has been neglected as it only changes the norm of the vector $\bm{\delta\phi}^{(m)}$, which is found later in the algorithm. We can restate the problem of finding $\bm{\delta \phi}$ as finding a linear combination of vectors. First, the geodesic $\Gamma^{(m)}$ and effective generators $\Omega_l(\bphi^{(m)})$ are written as vectors in the Lie algebra
\begin{align}
    \bm{\gamma}^{(m)} &= \sum_{j=1}^{N^2-1} \Tr{G_j \Gamma^{(m)}} \hat{e}_j, \\
    \bm{\omega}^{(m)}_l &= \sum_{j=1}^{N^2-1} \Tr{G_j \Omega_l(\bphi^{(m)})} \hat{e}_j,
\end{align}
where $\hat{e}_j$ are basis vectors. We now have a convex optimisation problem known as \emph{linear least-squares}~\cite[Section~1.2.2]{boyd2004convex}, with no constraints,
\begin{mini}|l|
{\bm{\delta\phi}^{(m)}}{\Bigg\Vert \sum_{j \forall G_j \in \mathcal{H}} \left[\delta\phi_j^{(m)} \bm{\omega}^{(m)}_j\right] - \bm{\gamma}^{(m)}\Bigg\Vert^2,}{}{}
\label{eq:convex_optimisation}
\end{mini}
where $\Vert \cdot \Vert$ is again the Euclidean 2-norm. Note that many terms in $\bm{\delta\phi}^{(m)}$ are constrained to be 0 because they do not appear in the restriction set $\mathcal{H}$. These constraints do not formally have to appear in the convex optimisation problem because they are not included in the summation. The number of parameters to optimise over is therefore $\vert \mathcal{H}\vert$, the cardinality of the set $\mathcal{H}$. 

Throughout the algorithm, an additional constraint is applied to the restricted parameter vectors $\bphi^{(m)}$. The parameter vectors are constrained to satisfy the commutator condition of Eq.~\eqref{eq:commutator_condition}, which is
\begin{align}
    [H(\bphi^{(m)}), \log(V)] = 0,
\end{align}
for all $m$. This constraint reduces the number of independent $\bphi$ parameter components, thereby decreasing the number of steps to reach a parameter vector that generates a unitary evolution sufficiently close to the target. Note that it depends on both $V$, and the restricting $\mathcal{H}$ whether commuting parameter constraints can be found. Additionally, since the Hilbert space increases exponentially with the number of qubits, the conditions from this constraint become increasingly difficult to compute. For a low number of qubits, such as three or four, it is straightforward to obtain with any symbolic programming language.

A golden section line search on the magnitude of the update vector $\delta^{(m)}$ can then be performed to maximise the update such that the metric $\Tr{U(\bphi^{(m+1)})^\dagger V}$ is largest \cite{kiefer1953sequential}. However, it is possible to reach local minima, where $\Tr{U(\bphi^{(m+1)})^\dagger V} \leq \Tr{U(\bphi^{(m)})^\dagger V}$. In this case, we generate a random vector $\bm{r}$, apply the restriction $\bm{r}_\mathcal{H} = R_\mathcal{H}\bm{r}$, and then use the Gram-Schmidt procedure to ensure the vector is orthogonal to the geodesic
\begin{align}
    \bm{\delta\phi}^{(m)} = \bm{r}_\mathcal{H} - \frac{\bm{r}_\mathcal{H}\cdot\bm{\gamma}^{(m)}}{\Vert\bm{\gamma}^{(m)}\Vert^2}\bm{\gamma}^{(m)}.
\end{align}
The Gram-Schmidt procedure reduces the chance that the subsequent steps of the algorithm find the same local minima. The steps of the algorithm are depicted in Fig.~\ref{fig:strategy}. We can control the strength with which we attempt to escape local minima by multiplying the resulting $\bm{\delta\phi}^{(m)}$ with a scalar $\eta$, which is one of the hyperparameters of our algorithm. 
Finally, we terminate the optimisation when the infidelity is smaller than some desired precision $\varepsilon$. The full algorithm is given in Algorithm~\ref{alg:unitary_gate_design}.
\begin{algorithm}
\caption{Geodesic gate design.}\label{alg:unitary_gate_design}
\KwIn{$V$, $\btheta, R_{\mathcal{H}},  \varepsilon, \eta$}
\KwOut{$\bphi$}
\text{Obtain the Jacobian function:}\\
\For{$l \in (1,\ldots, N^2-1$)}{
$dU_l(\bx) = \partial_{x_l} \mathfrak{Re}[U(\bx)] + i \partial_{x_l}\mathfrak{Im}[U(\bx)]$}

\text{Define a fidelity function:} $F(\btheta, V) = \Tr{U(\btheta)^\dagger V}/N$

\text{Perform the Hamiltonian restriction:} $\bphi \gets R_{\mathcal{H}}\btheta$

Solve the commutator condition reducing the number of independent $\bphi$ components: $[H(\bphi), \log(V)] = 0$

\text{Update $\bphi$ respecting the commutator condition:}\\
\While{$F(\bphi, V) < 1-\varepsilon$}{
    \text{Compute the effective generators at $\bphi$:}\\
    \For{$l \in (1,\ldots, N^2-1)$}{$\Omega_l(\bphi) \gets U^\dag(\bphi) dU_l(\bx)|_{\bphi}$
    
    $\bm{\omega}_l \gets \sum_{j=1}^{N^2 -1} \Tr{G_j \Omega_l(\bphi)} \hat{e}_j$}
    
    \text{Find the geodesic:}\\
    $\Gamma \gets -i \log(U(\bphi)^\dagger V)$\\
    $\bm{\gamma} \gets \sum_{j=1}^{N^2 -1} \Tr{G_j \Gamma(\bphi)} \hat{e}_j$ 

    \FuncSty{Minimise}$\left(\Big\Vert \sum_{j \forall G_j \in \mathcal{H}} \left[ \delta\phi_j \bm{\omega}_j \right] - \bm{\gamma}\Big\Vert^2\right) \textrm{ for } \bm{\delta\phi}$

    \FuncSty{Maximise}$\left(F(\bphi + \delta \bm{\delta\phi}, V)\right) \textrm{ for } \delta$
    
    \uIf{$F(\bphi + \delta\bm{\delta\phi}, V) < F(\bphi, V)$}{choose random $\bm{r}$

    $\bm{r}_\mathcal{H} \gets R_\mathcal{H}\bm{r}$
    
    $\bm{\delta\phi} \gets \bm{r}_\mathcal{H} - \frac{\bm{r}_\mathcal{H}\cdot\bm{\gamma}}{\Vert\bm{\gamma} \Vert^2} \bm{\gamma}$
    
    $\bphi \gets \bphi + \eta \bm{\delta\phi}$
    }
    \Else{
    $\bphi \gets \bphi + \delta \bm{\delta\phi}$
    }

}
\end{algorithm}

\section{\label{sec:numerics}Numerical experiments}
To test the validity of our approach, we construct several quantum gates with the Hamiltonian restriction set $\mathcal{H}_{2-\mathrm{local}}$, which is the set of Pauli words that have only one- and two-body interactions. We see that $\vert\mathcal{H}_{2-\mathrm{local}}\vert = 9n(n+1)/2 + 3n = O(\log(N)^2)$. For this restriction, an optimal solution vector $\bm{\delta\phi}^{(m)}$ can be found for the least-squares problem efficiently in $N$ -- in time that scales as $O(\log(N)^2 N^4)$. 

Throughout the rest of this section, we initialise the parameters as $\bphi^{(0)}\sim \mathcal{U}(-1,1)$. We set $\eta = 1$, which gives a reasonable trade off between exploration and exploitation for the Gram-Schmidt restarting subroutine of our algorithm.

\subsection{Toffoli and Fredkin gate}
\begin{figure}[htb!]
    \centering
    \subfloat[Toffoli]{\includegraphics[scale=0.44]{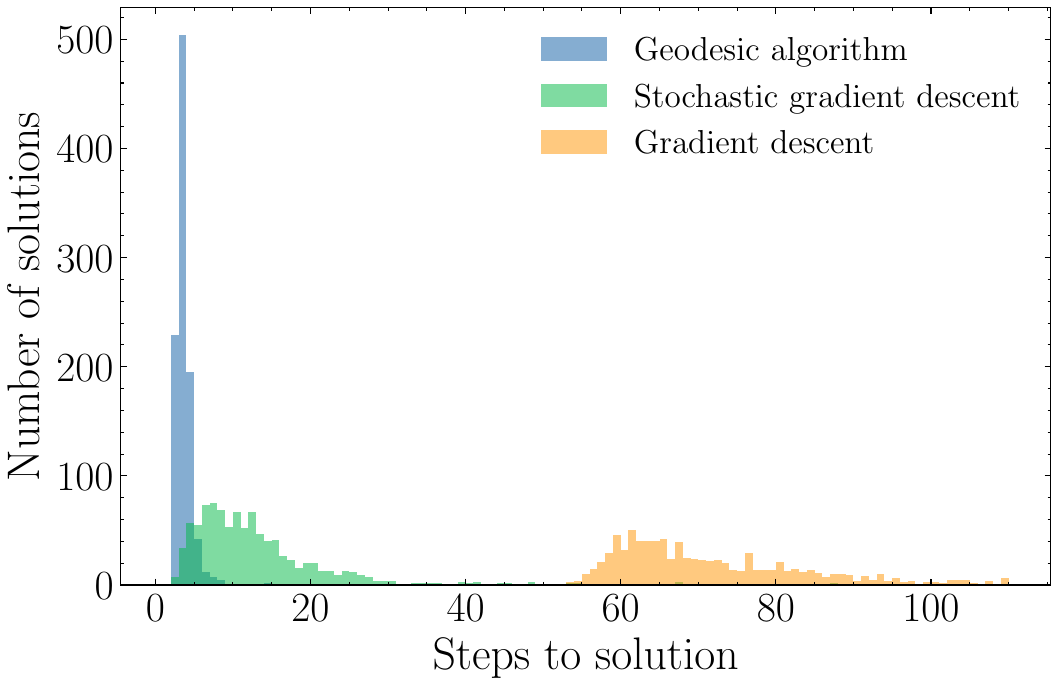}}

    \subfloat[Fredkin]{\includegraphics[scale=0.44]{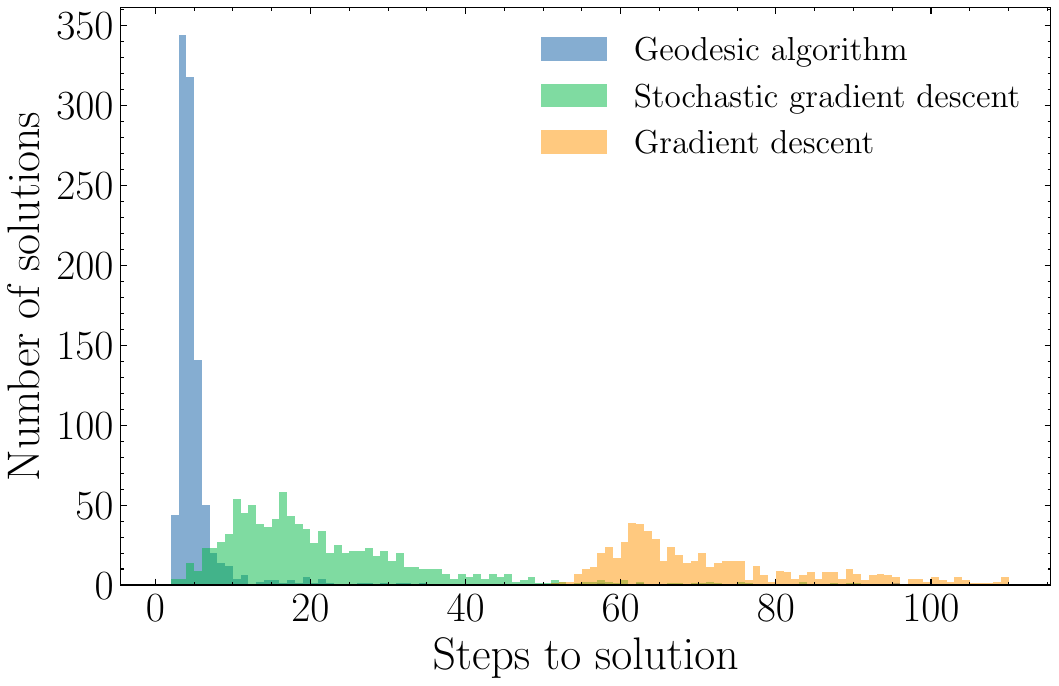}}
    \caption{Comparison of the number of steps required before a solution with infidelity less than $\varepsilon=0.001$ is found for: our geodesic algorithm (blue), stochastic gradient descent (SGD) of Ref.~\cite{Innocenti2020} (green), and a simple gradient descent optimiser that uses gradients computed with JAX (orange). For all methods, we consider $1000$ random initialisations of the parameters. (a) For the Toffoli gate, both the geodesic algorithm and SGD algorithm find a solution 100\% of the time, whereas the JAX gradient descent algorithm finds a solution 98\% of the time. (b) For the Fredkin gate, the geodesic algorithm finds a solution 99.3\% of the time, the SGD finds a solution 100\% of the time. The JAX gradient descent algorithm struggles to find a solution for this gate, with a success rate of only 71.5\%.}
    \label{fig:toffoli_fredkin_comparison}
\end{figure}

We first consider two gates that can be combined with single qubit unitaries to form a universal gate set, the Toffoli and Fredkin gate \cite{nielsen_quantum_2006}. As a benchmark, we consider the work of Innocenti et al. \cite{Innocenti2018}, which attempts to learn a unitary generated by a time-independent Hamiltonian from a set of training data states (see App.~\ref{app:innocenti}). Additionally, we compare with a simple gradient descent scheme, JAX gradient descent (see App. \ref{app:jaxgd}). For both the Toffoli and the Fredkin gate, it is possible to further constrain the parameters $\bphi$ to always ensure that the restricted Hamiltonian commutes with the target. We use the symbolic mathematical package Sympy to find these constraints \cite{2017sympy}. In Fig.~\ref{fig:toffoli_fredkin_comparison}, we show a histogram of the number of steps required to find a high precision Toffoli and Fredkin gate for 1000 random initialisations. Our geodesic algorithm requires significantly fewer steps on average than the other methods. We provide the data to reproduce these figures at~\cite{our_data} and give the parameters for three example solutions for the Toffoli and Fredkin in App.~\ref{app:parameters}.

\subsection{Weight-$k$ Parity checks}
\begin{figure}[htb!]
    \centering
    \subfloat[Weight-$k$ $Z$ Parity ($\mathrm{WZ}_k$)]{\includegraphics[scale=0.47]{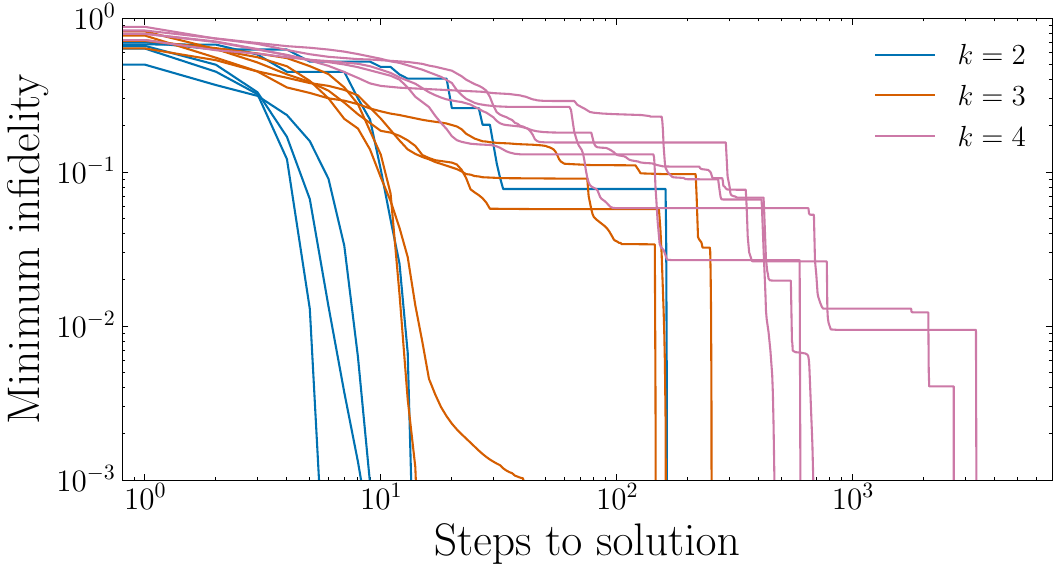}}

    \subfloat[Weight-$k$ $X$ Parity ($\mathrm{WX}_k$)]{\includegraphics[scale=0.47]{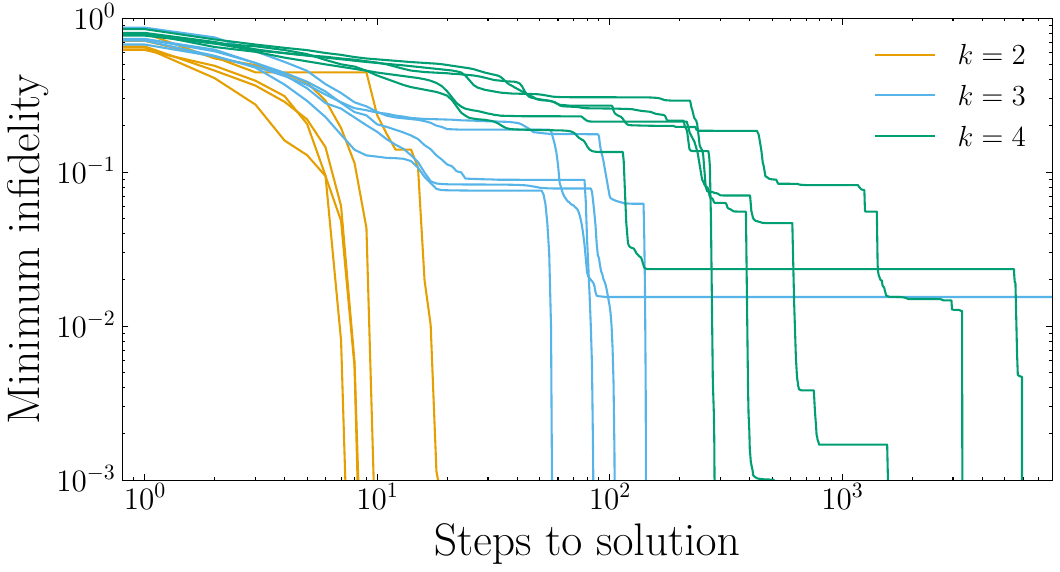}} 
    \caption{Plot of the minimum infidelity $1 -  F(\btheta, V)$ up to each step in the algorithm for weight-$k$ parity checks with $k=\{2,3,4\}$. We show 5 optimisation runs. The steps are shown until the target precision of $0.001$ infidelity is reached. In subsequent steps significantly lower infidelities can be obtained for some trajectories. We do not use the commuting ansatz of Eq.~\eqref{eq:commutator_condition} here due to the excessive amount of time it takes to compute the ansatz for larger gates.}
    \label{fig:parity_checks_trajectories}
\end{figure}

In order to build fault-tolerant quantum computers, we will need robust error correction to mitigate the effect of noise in current quantum devices. A widely-used family of quantum error correction codes are so-called stabilizer codes, which rely on measuring syndromes and using the resulting measurement outcome to correct a computation. Of particular interest are surface codes~\cite{fowler_surface_2012}, and the toric code \cite{bravyi_quantum_1998}, which rely on stabilizers that are geometrically local on a two dimensional grid. Crucial to these codes are so called parity checks; circuits that can detect flipped bits or phase changes. Here, we consider the unitary corresponding to a weighted $X$ and weighted $Z$ parity check of the form
\begin{align}
    \mathrm{WZ}_k &= \frac{1}{2}(I^{\otimes k} + Z^{\otimes( k-1)} I + I^{\otimes( k-1)} X - Z^{\otimes( k-1)} X) \nonumber\\
    \mathrm{WX}_k &= \frac{1}{2}(I^{\otimes k} + X^{\otimes(k-1)} I + I^{\otimes(k-1)} X - X^{\otimes k}),
\end{align}
respectively. Here $k$ denotes the size of the check. 
In Fig.~\ref{fig:parity_checks_trajectories}, we see that our algorithm is capable of finding accurate solutions for the weight-$k$ parity check unitary, for $k=2,3,4$, which correspond to gates comprising 3, 4, and 5 qubits respectively. Additionally, we found a solution for $k=5$, which is a 6 qubit gate, after approximately $3 \times 10^4$ steps. Some trajectories in Fig.~\ref{fig:parity_checks_trajectories} show the optimisation plateauing for a number of steps, either temporarily or in one case until the maximum step number is reached. In these cases, the optimisation can repeatedly find a solution that would require additional Hamiltonian terms, which are not in the restriction (e.g. not physical), to reach a high fidelity. The Gram-Schmidt procedure provides a method to avoid these infidelity plateaus, however, at the cost of tuning the hyperparameter $\eta$.

\section{Physical Restriction\label{sec:physical_restriction}}
In the previous section, the Hamiltonian restriction considered was all one- and two-body interactions, $\mathcal{H}_\textrm{2-local}$. Here, we consider a physically more relevant restriction, composed of the Hamiltonian terms of an anisotropic Heisenberg model $\mathcal{H}_\textrm{Heis}$, with Hamiltonian
\begin{multline}
    H_\textrm{Heis} = \sum_{i,j=1}^{n} J_{ij}^{x} \sigma^{x}_i \sigma^{x}_j + J_{ij}^{y} \sigma^{y}_i \sigma^{y}_j + J_{ij}^{z} \sigma^{z}_i \sigma^{z}_j \\ + \sum_{i=1}^{n} h_i^x  \sigma^x_i + h_i^y  \sigma^y_i + h^z_i \sigma^z_i.
    \label{eq:heisenberg_hamiltonian}
\end{multline}
The Hamiltonian is a general quantum simulator in which many typical Hamiltonian models can be realised: the Ising model, Ising model with transverse field, XY model, and the Heisenberg model~\cite{mbeng_quantum_2024}. The model is therefore of practical interest for general quantum simulators~\cite{Altman2019}, and is physically plausible to engineer in ion traps~\cite{porras_effective_2004,monroe_programmable_2021} or Rydberg atom systems~\cite{morgado_quantum_2021}. 

The restricted parameter vector $\bphi$ that we find is therefore a vector of $J_{ij}^{\alpha}$ and $h_i^{\alpha}$ for $\alpha= x,y,z$. For the Toffoli and Fredkin gate, solutions are found even with this more severe restriction in 10 to 100 steps. Example solutions are given in App.~\ref{app:heisenberg_restriction}.

The key advantage of these gates is that faster circuit run times are possible compared to using one- and two-qubit gates. Here we take the specific case of the Toffoli gate. Considering a one- and two-qubit universal gate set of a CNOT gate and local gates, the Toffoli gate can be optimally implemented with six CNOT gates, two Hadamard gates, four T gates, and three $\textrm{T}^\dagger$ gates~\cite{shende_cnot-cost_2009} -- the circuit diagram is given in Fig.~\ref{fig:toffoli_circuit_decomp} in App.~\ref{app:heisenberg_restriction}.

In order to perform a time comparison, we make the simplifying assumption that the magnitude of each Hamiltonian term, both the one- and two-local terms, is individually bounded. For the purposes of comparison, we rescale the largest term of the Hamiltonian to 1. The time for the Hamiltonian dynamics is therefore given by the magnitude of the largest $\bphi$ parameter in the solution.

The Hamiltonian $H_\textrm{CNOT}$ that implements the CNOT gate is found by taking the principle branch of the logarithm,
\begin{align}
    H_\textrm{CNOT} = \frac{\pi}{4} (\sigma_1^z \sigma_2^x - \sigma_1^z - \sigma_2^x ),
\end{align}
where we have neglected global phase in the CNOT gate and take the unitary definition of Eq.~\eqref{eq:U_theta} for consistency.
We note that this Hamiltonian is not actually possible in the Heisenberg restriction $\mathcal{H}_\textrm{Heis}$, but is possible with the restriction $\mathcal{H}_\textrm{2-local}$. The principle branch gives a lower bound for the time to implement the CNOT gate, $t_\textrm{CNOT} = \pi/4$. The Hadamard gate is implemented by the Hamiltonian $H_\textrm{Hadamard} = -\frac{\pi}{2\sqrt{2}}(\sigma^{x}_1 + \sigma^{z}_1)$, which gives a time of $t_\textrm{Hadamard} = \pi/2\sqrt{2}$. Similarly, $H_\textrm{T} = \frac{\pi}{8} \sigma^z_1$ implements the T gate, giving a time $t_\textrm{T} = \pi/8$, which is the same time for the $\textrm{T}^\dagger$ gate. Since some of the gates can be performed simultaneously, the total time for the Toffoli to be performed by decomposition with the aforementioned universal gate set is $t_\textrm{decomp} = 2 t_\textrm{Hadamard} + 5 t_\textrm{CNOT} + 4 t_\textrm{T}$. Table~\ref{table:circuit_times} compares the time to implement a Toffoli gate using the decomposed circuit with the time for a direct implementation using the result of the geodesic algorithm. In both cases, we consider the restriction $\mathcal{H}_\textrm{2-local}$ and the restriction $\mathcal{H}_\textrm{Heis}$ of anisotropic Heisenberg interactions. We also note again how for the decomposition the time for the CNOT would be more with an equivalent Heisenberg restriction. With this restriction, using the geodesic algorithm, we find a CNOT time of $t_\textrm{CNOT} = 2.36$. 

We also note that for the Fredkin gate, as shown in App~\ref{app:heisenberg_restriction}, we find a time $1.17$ for the direct implementation with the 2-local restriction set $\mathcal{H}_\textrm{2-local}$ and $2.08$ for the Heisenberg restriction set $\mathcal{H}_\textrm{Heis}$. Circuit decompositions with a two qubit gate, as with the Toffoli gate shown in Table~\ref{table:circuit_times}, would lead to significantly longer circuit run times for the Fredkin gate as well.  

\begin{table}
\begingroup
\setlength{\tabcolsep}{7pt} 
\renewcommand{\arraystretch}{1.2} 
\begin{tabular}{|l|l|l|c|}
\hline
Gate    & Implementation                                & Restriction            & Time  \\ \hhline{|=|=|=|=|}
T / $\textrm{T}^\dagger$ & -        & -                &  0.39 \\ \hline
Hadamard & -        & -                & 1.11  \\ \hline

CNOT & -        & $\mathcal{H_\textrm{2-local}}$                & 0.79  \\  \hline
CNOT & -        & $\mathcal{H_\textrm{Heis}}$               & 2.36   \\ \hline
Toffoli & Decomposition        & $\mathcal{H_\textrm{2-local}}$                & 7.72  \\ \hline
Toffoli & Decomposition & $\mathcal{H}_\textrm{Heis}$ & 15.59 \\ \hline
Toffoli & Geodesic algorithm                               & $\mathcal{H_\textrm{2-local}}$                & 1.30  \\ \hline
Toffoli & Geodesic algorithm                                & $\mathcal{H}_\textrm{Heis}$ & 2.58  \\ \hline
\end{tabular}
\endgroup
\caption{The time to implement gates using a circuit decomposition with CNOT gates is compared to a direct implementation found using the geodesic algorithm. The circuit decomposition for the Toffoli is given in App.~\ref{fig:toffoli_circuit_decomp} in App.~\ref{app:heisenberg_restriction}. We consider both the 2-local restriction $\mathcal{H_\textrm{2-local}}$ and the anisoptric Heisenberg restriction $\mathcal{H}_\textrm{Heis}$. The time for the Toffoli gate uses the minimum time for the CNOT gate with the Heisenberg restriction $\mathcal{H}_\textrm{Heis}$ for the circuit decomposition and the minimum time for the direct implementation of the Toffoli gate. The minimum time is found by taking the minimum for 1000 instances of the geodesic algorithm. App.~\ref{app:heisenberg_restriction} shows the parameters that give the minimal time for implementing the CNOT gate and Toffoli gate with restrictions. }
\label{table:circuit_times}
\end{table}

\section{\label{sec:discussion}Discussion}
The commuting ansatz of Eq.~\eqref{eq:commutator_condition} is the requirement that the Hamiltonian at each step $H(\bphi^{(m)})$ must commute with the principal logarithm of the target $\log(V)$. For a low number of qubits, three or four, we find that the ansatz offers a considerable advantage in the efficiency to discover the solution. Whereas for five or more qubits we find that the solution is often discovered more slowly than by simply ignoring the commuting constraint. The severe restriction imposed by the commuting ansatz on the allowed updates of the parameters prevents the geodesic algorithm from more directly moving towards the target unitary on the manifold. The commutator condition is of practical importance since it works well for three to five qubits. As the number of qubits increases, restricted Hamiltonian sets with 2-local Hamiltonians would generate a vanishingly small proportion of possible quantum gates. This is because the number of 2-local Hamiltonian terms grows polynomially with the number of qubits as $O(n^2)$ but the full Lie algebra dimension grows exponentially as $O(2^n)$. It is therefore unlikely that many complex multi-qubit quantum gates with more than six qubits are possible with Hamiltonians that arise in current hardware platforms. In fact, even for three qubits, there will be gates that cannot be generated with a single time-independent Hamiltonian since the Lie algebra vector space with the restriction does not span the full Lie algebra vector space. Whether a gate is possible with the restricted Hamiltonian would also depend on the available Hamiltonian terms in the restriction. Three and four qubit gates would already lead to significant circuit simplifications. This is because a deep circuit will generically have a large number of possible multi-qubit gates that can be found with our geodesic algorithm. The time reduction increases additively with circuit depth. 

We find that our geodesic algorithm is significantly more efficient than stochastic gradient descent and gradient descent algorithms for finding a restricted generating Hamiltonian of a desired unitary gate of even up to 6 qubits. 
Gradient descent methods also quickly become infeasible for larger quantum gates, and have not been able to find useful quantum gates of more than 3 qubits. 
Our geodesic algorithm takes advantage of how the problem can be described on the Riemannian manifold of $\SU{N}$. We use the fact that the location of the solution on the manifold, the target unitary $V$, is known in advance. Typically, in machine learning, the location of the solution in the phase space is unknown. In these cases, the computation and minimisation of a cost function is required and gradient descent algorithms cannot use information about the location of the target point. In contrast, our geodesic algorithm is aware of the target point and attempts to push the parameters towards that point on the manifold at each step. The effective generators $\Omega_l$ provide the necessary information for how the Hamiltonian parameters should be updated in order for the next set of parameters to follow the path of the geodesic to the solution as closely as possible. 

Time-independent Hamiltonians generate geodesic curves on the manifold since they are defined by a one-parameter subgroup of $\SU{N}$ parameterised by $t$. These geodesics start at the identity element at $t=0$ and reach the desired unitary evolution at $t=1$. In general, due to the Hamiltonian strengths, these geodesic curves are not necessarily the shortest paths along the manifold, which for a given restriction could be a time-dependent Hamiltonian~\cite{dowling_geometry_2008}. Consider a restriction $\mathcal{H}$ and the additional Pauli words $\mathcal{Q} = \mathcal{P} \setminus \mathcal{H}$. The principal generator of a unitary gate generally contains terms from $\mathcal{H}$ and $\mathcal{Q}$. We can define a special commuting case where the specific terms of the principal generator in $\mathcal{Q}$ commute with those in $\mathcal{H}$. In this case, the minimal path along the manifold to generate the desired unitary (given the restriction $\mathcal{H}$) is actually a geodesic generated by a time-independent Hamiltonian~\cite[Section IV.C.1]{dowling_geometry_2008}. Crucially, the minimal path for the restriction $\mathcal{H}$ is neither a quantum circuit with multiple layers of gates nor a time-dependent Hamiltonian. Thus, for an initialisation of parameters close to the identity (such that the parameters start small in magnitude), our geodesic algorithm finds the most efficient implementation possible of parity check gates with only one- and two-local Hamiltonian terms.

\section{\label{sec:conclusion}Conclusion}
Hamiltonians underpin quantum computation. In this work, we have introduced a geodesic algorithm to find time-independent Hamiltonians that implement multi-qubit unitaries. 
First, we considered the simplest `physical' restriction, using all one- and two-local Hamiltonian terms, but we show that our algorithm can be applied to more constrained Hamiltonians, for example, restricting to anisotropic Heisenberg Hamiltonian terms. We demonstrated a significant run time advantage in this case. In particular, for the Toffoli gate, we found a run time improvement of a factor of six by using a single time-independent Hamiltonian with engineered couplings from our geodesic algorithm.

Current experimental platforms for quantum computing offer more complex Hamiltonians than are typically utilised for quantum gates. In fact, the gates implemented at the hardware level are often at most only two-qubit gates. Our geodesic algorithm opens up the possibility of finding time-independent Hamiltonian couplings so that larger, more complex quantum gates can be implemented directly. Additionally, we could combine multiple blocks of time-independent quantum gates together to form more complex circuits, where we find the required Hamiltonian couplings of each individual block using our geodesic algorithm. Not only could this lead to less noisy gates, it could also reduce the total time to run a circuit on the hardware, as we have demonstrated for the simple example of a Toffoli gate. This is crucial for NISQ applications where we have a limited coherence time and gives the significant advantage of increasing the clock speed for fault-tolerant quantum computation by reducing the parity check times.
In general, focusing on experimentally generating more complex Hamiltonians for quantum computing could be a more fruitful approach for decreasing circuit depth than more traditional compilation methods.

\section{Acknowledgements}	
DL acknowledges support from the EPSRC Centre for Doctoral Training in Delivering Quantum Technologies, grant ref.~EP/S021582/1. JC acknowledges support from the Natural Sciences and Engineering Research Council (NSERC), the Shared Hierarchical Academic Research Computing Network (SHARCNET), Compute Canada, and the Canadian Institute for Advanced Research (CIFAR) AI chair program. Resources used in preparing this research were provided, in part, by the Province of Ontario, the Government of Canada through CIFAR, and companies sponsoring the Vector Institute \url{www.vectorinstitute.ai/#partners}.

\section{Competing interests}
DL, RW, and SB declare a relevant patent application: United Kingdom Patent Application No.~2400342.8.

\onecolumngrid
\renewcommand{\thesection}{\Alph{section}}
\renewcommand{\theequation}{\Alph{section}\arabic{equation}}
\renewcommand{\thefigure}{\Alph{section}\arabic{figure}}
\renewcommand{\thetable}{\Alph{section}\arabic{table}}
\renewcommand{\thesubsection}{\Roman{subsection}} 
\counterwithin*{equation}{section}
\counterwithin*{table}{section}
\counterwithin*{figure}{section}
\setcounter{section}{0}
\pagebreak
\begin{center}
\textbf{\large Appendix}
\end{center}
\setcounter{equation}{0}
\setcounter{figure}{0}
\setcounter{table}{0}
\makeatletter
\section{Stochastic Gradient Descent\label{app:innocenti}}

Here, we summarise the stochastic gradient descent algorithm from Ref.~\cite{Innocenti2018} which is used for the numerical results in the main text. We consider the same setup as in the main text, with $\mathcal{H}\subset \mathcal{P}$ the restricted set of Pauli strings and $\bphi = R_{\mathcal{H}} \btheta$ the parameters of the restricted Hamiltonian. Instead of considering the direct minimisation of the unitary infidelity of Eq.~\eqref{eq:infid}, we can consider the average state infidelity over a set of $d$ states $\mathcal{D}=\{\ket{\psi_i}\}^d$,
\begin{align}
    \mathcal{I}_{\mathcal{D}}(\bphi) = 1 - \frac{1}{d}\sum_i^d
    \abs{\bra{\psi_i}U^\dag(\bphi) V \ket{\psi_i}}^2 \label{eq:infidelity_train}
\end{align}
Minimizing the function $\mathcal{I}_{\mathcal{D}}(\bphi)$ for some finite set of states is an example of supervised learning~\cite{murphy2022ml}. Note that if $\mathcal{D}=\{\ket{b_i}\}^{N}$ is the set of all computational basis states, where $\ket{b_i}$ is a computational basis state corresponding to bit $i$, then $\mathcal{I}_{\mathcal{D}}(\bphi) \equiv \mathcal{I}(\bphi)$. The cost of evaluating $\mathcal{I}_{\mathcal{D}}(\bphi)$ is $O(d\times N)$ compared to $O(N^2)$ for $\mathcal{I}(\bphi)$, hence it is a cheaper method of estimating the infidelity. Additionally, the stochasticity of the optimisation could be useful in escaping from local minima.

To minimise the cost in Eq.~\eqref{eq:infidelity_train} in practice one can perform a stochastic gradient descent algorithm, which uses the gradient over randomly sampled elements of a data set to minimise a cost function. We randomly initialise the parameters of our model $\bphi^{(0)}\sim \mathcal{U}(-1,1)$. At the $m$th step of the optimisation, we create a data set $\mathcal{D}^{(m)}=\{\ket{\psi_i^{(m)}}\}^{d_{\mathrm{train}}}$ that is randomly sampled from the Haar measure. Then, we obtain the gradients with respect to $\bphi$ using the automatic differentiation framework Theano \cite{al2016theano}. We update the parameters of the model with a standard gradient descent update step with an initial learning rate $\lambda^{(0)}=1$. Throughout the optimisation, the learning rate is decayed as
\begin{align}
    \lambda^{(m+1)} = \frac{\lambda^{(0)}}{1 + \kappa \times m}
\end{align}
at each step $m$, where $\kappa$ is defined as the gradient decay rate.
In the spirit of machine learning, we keep a fixed validation set of states $\mathcal{D}_{\mathrm{val}}=\{\ket{\psi_i^{(m)}}\}^{d_{\mathrm{test}}}$ to track the performance of the model during the optimisation. If the test set is large enough, this should give an accurate estimate of the true infidelity.
For the experiments in the main text, we make use of the hyperparameters in Table~\ref{tab:hyps_innocenti} which are are the same as in Ref.~\cite{Innocenti2018}.
\begin{table}[htb!]
    \centering
    \begin{tabular}{c|c}
        Name & Value\\\hline\hline
        ${d_{\mathrm{train}}}$ & 200  \\\hline
        ${d_{\mathrm{test}}}$ & 100  \\\hline
        Gradient decay rate $\kappa$ & 0.005 \\\hline
    \end{tabular}
    \caption{Stochastic gradient descent hyperparameters}
    \label{tab:hyps_innocenti}
\end{table}

\section{Gradient Descent \label{app:jaxgd}}
Here we summarise the gradient descent algorithm which is used for the numerical results in the main text. A simple algorithm to find the generator of a target unitary $V$ can be described as follows. We consider the same setup as in the main text, with $\mathcal{H}\subset \mathcal{P}$ the restricted set of Pauli strings and $\bphi = R_{\mathcal{H}} \btheta$ the parameters of the restricted Hamiltonian. We consider the minimisation of the infidelity in Eq.~\eqref{eq:infid} and attempt to solve this optimisation problem via gradient descent. At each step $m$ of the optimisation, we obtain the gradients
\begin{align}
    \frac{\partial U(\bphi^{(m)})}{\partial \phi_j}\bigg\vert_{\bphi^{(m)}} = U(\bphi) \Omega_j(\bphi)
\end{align}
via automatic differentiation with Jax \cite{jax2018github}, since the matrix exponential \texttt{jax.scipy.linalg.expm} is supported as a differentiable function. Once we obtain the gradients, we use ADAM \cite{kingma2015adam} to update the parameters $\bphi$. For the results in the main text we used a learning rate of $\lambda=10^{-1}$. We do not include the commutator condition of Eq.~\eqref{eq:commutator_condition}, nor do we use a line search to find the optimal step size.
\clearpage
\section{\label{app:parameters}Parameters}
In this section we show three sets of parameters for the Toffoli and Fredkin gate found by our algorithm for the restriction $\mathcal{H}_\textrm{2-local}$. For both gates, we initialise the parameters close to the identity, which in practice helps with finding sparse solutions where many of the Hamiltonian terms have coefficients close to zero. All solutions shown here have an error of $\varepsilon<10^{-3}$. 

\subsection{Toffoli}
\begin{figure}[htb!]
    \centering
    \subfloat[Seed \#1]{\includegraphics[width=0.97\textwidth]{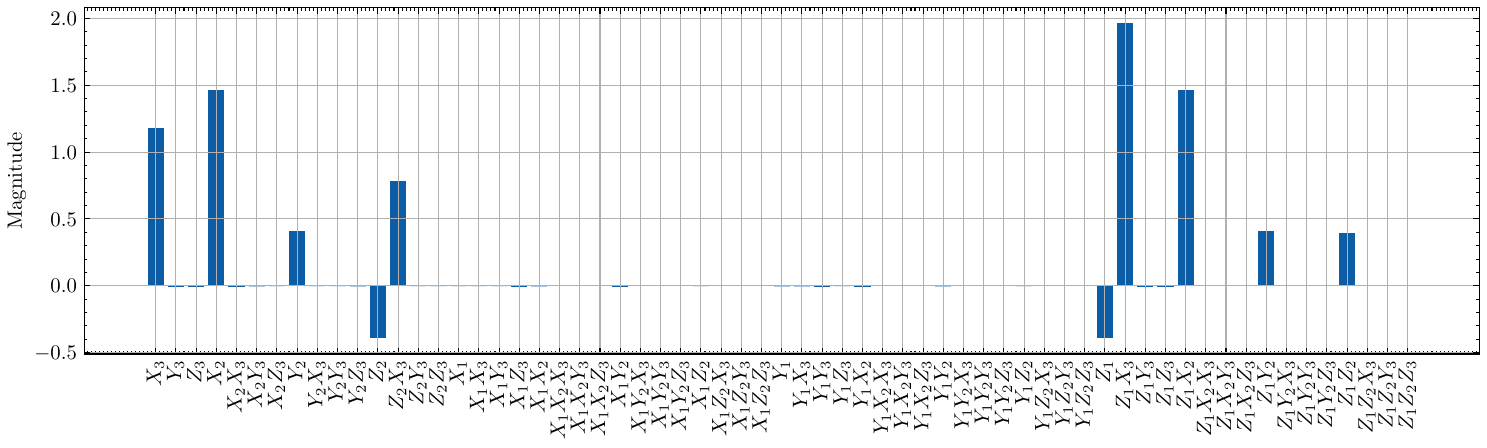}}
    
    \subfloat[Seed \#2]{\includegraphics[width=0.97\textwidth]{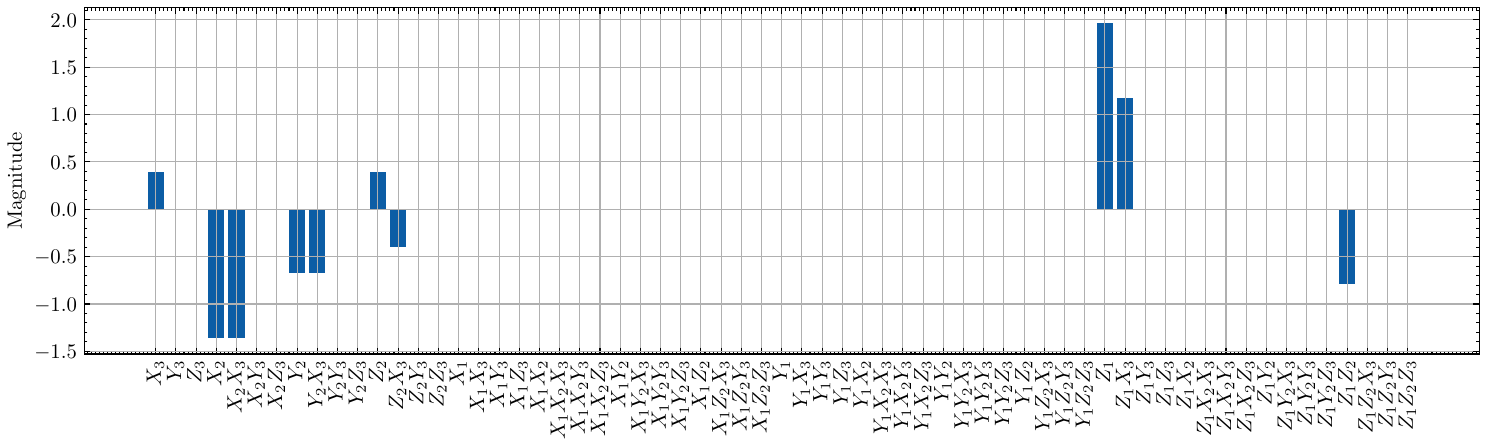}}
    
    \subfloat[Seed \#3]{\includegraphics[width=0.97\textwidth]{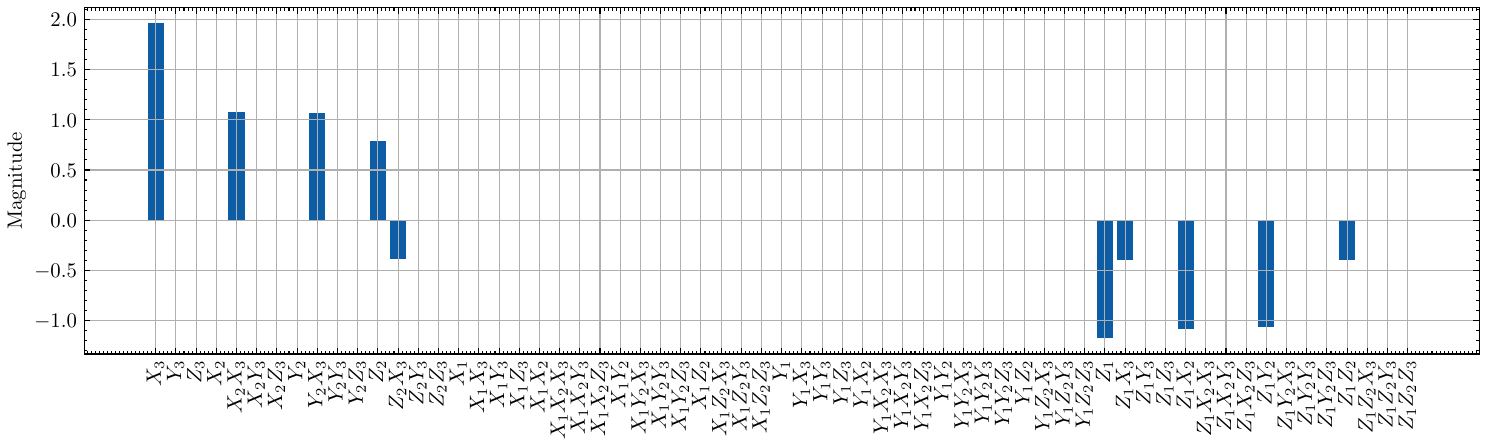}}

    \caption{Bar plot of parameter magnitudes for Toffoli gate. We show three different initialisations of initial parameters in (a), (b) and (c). Note how we find many solutions consisting mostly of local fields and a few two-body interactions.}
    \label{fig:toffoli_params}
\end{figure}
\clearpage
\subsection{Fredkin}
\begin{figure}[htb!]
    \centering
    \subfloat[Seed \#1]{\includegraphics[width=0.97\textwidth]{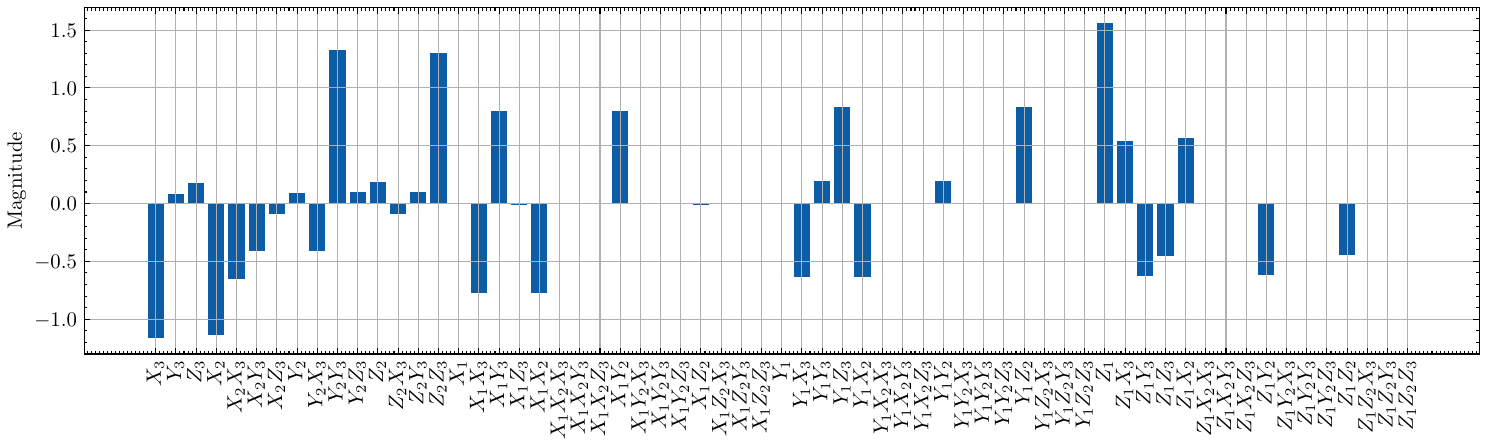}}
    
    \subfloat[Seed \#2]{\includegraphics[width=0.97\textwidth]{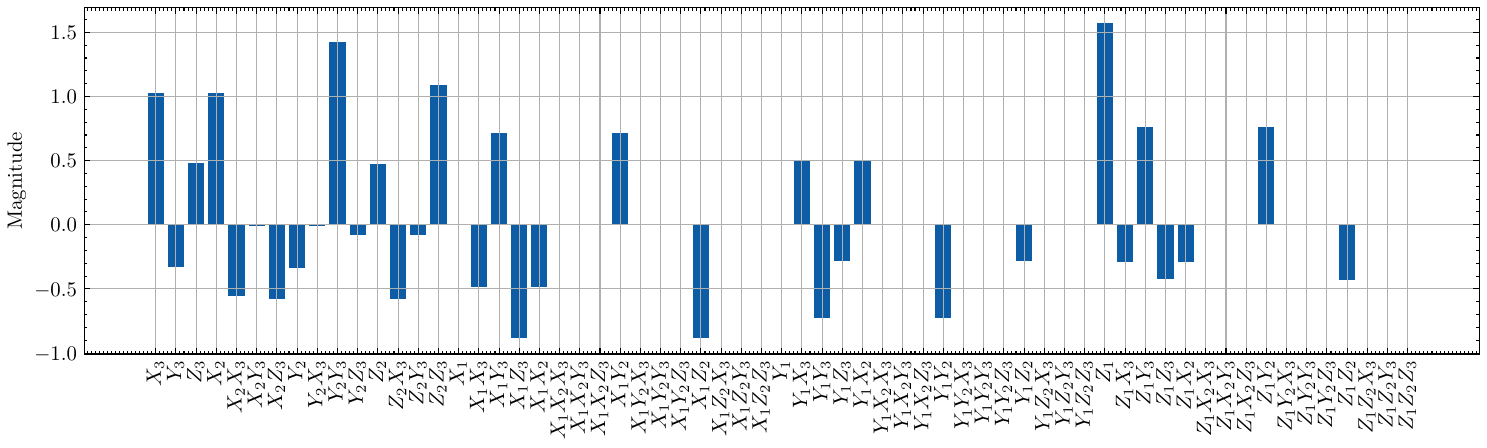}}
    
    \subfloat[Seed \#3]{\includegraphics[width=0.97\textwidth]{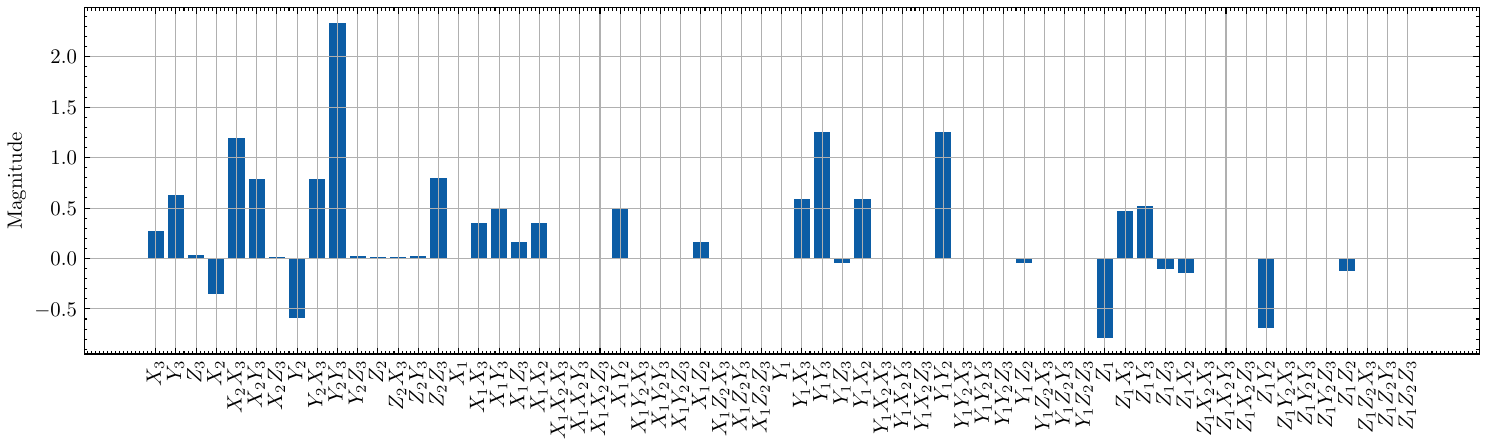}}

    \caption{Bar plot of parameter magnitudes for Fredkin gate. We show three different initialisations of initial parameters in (a), (b) and (c). The solutions are much less sparse than for the Toffoli, although some fields have small enough values that can be ignored without much loss of fidelity.}
    \label{fig:fredkin_params}
\end{figure}

\section{Physical Restriction\label{app:heisenberg_restriction}}
The Toffoli gate can be decomposed into CNOT gates and local gates by the circuit shown in Fig.~\ref{fig:toffoli_circuit_decomp}.
\begin{figure*}[h!]
    \centering
        \begin{quantikz}
        & \ctrl{1} & \ghost{T} &  \\
        & \ctrl{1} & \ghost{T^\dag} &  \\
        & \targ{} & \ghost{T^\dag} &  
    \end{quantikz}  =  \begin{quantikz}
        & \qw & \qw & \qw & \ctrl{2} & \qw & \qw & \qw & \ctrl{2} & \qw & \ctrl{1} & \gate{T} & \ctrl{1} & \qw \\
        & \qw & \ctrl{1} & \qw & \qw & \qw & \ctrl{1} & \qw & \qw & \gate{T} & \targ{} & \gate{T^\dag} & \targ{} & \qw \\
        & \gate{H} & \targ{} & \gate{T^\dag} & \targ{} & \gate{T} & \targ{} & \gate{T^\dag} & \targ{} & \gate{T} & \gate{H} & \qw &\qw & \qw 
    \end{quantikz}    
    \caption{Circuit decomposition of Toffoli gate into CNOT, T, and Hadamard gates. }
    \label{fig:toffoli_circuit_decomp}
\end{figure*}

The restriction set $\mathcal{H}_\textrm{2-local}$ contains all one- and two-body interaction terms. The Heisenberg restriction $\mathcal{H}_\textrm{Heis}$ includes only terms that appear in Eq.~\eqref{eq:heisenberg_hamiltonian} of the main text. The following sections show examples of gate solutions with the 2-local and Heisenberg restrictions. We show examples with the minimal run time (as defined in Section~\ref{sec:physical_restriction}) after running 1000 instances of the geodesic algorithm. These are the instances used for the times in Table~\ref{table:circuit_times}. 

\subsection{CNOT}
The minimal time CNOT gates found with 1000 instances of the geodesic algorithm.
\begin{figure}[htb!]
    \centering
    \subfloat[Restriction set $\mathcal{H}_\textrm{2-local}$]{\includegraphics[width=0.65\textwidth]{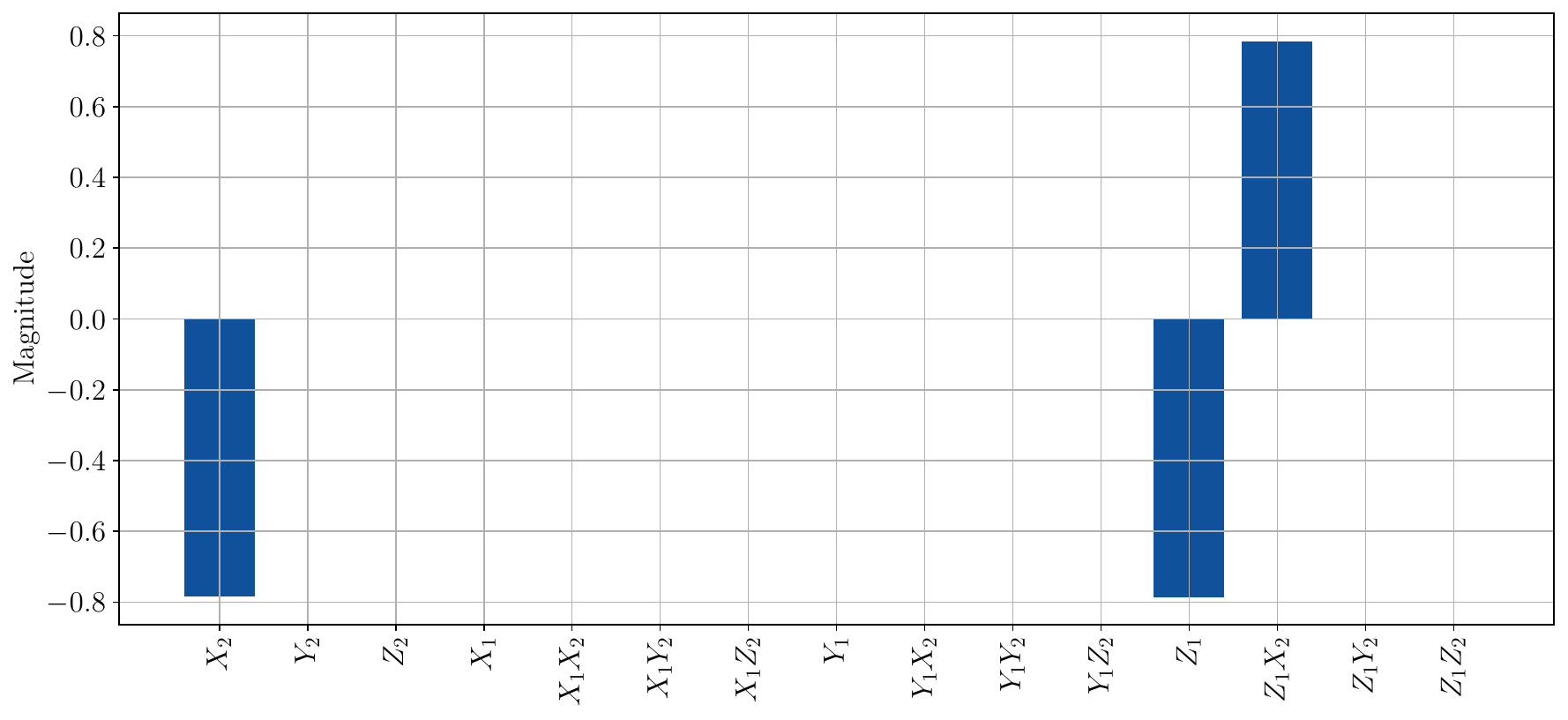}}
    
    \subfloat[Restriction set $\mathcal{H}_\textrm{Heis}$]{\includegraphics[width=0.65\textwidth]{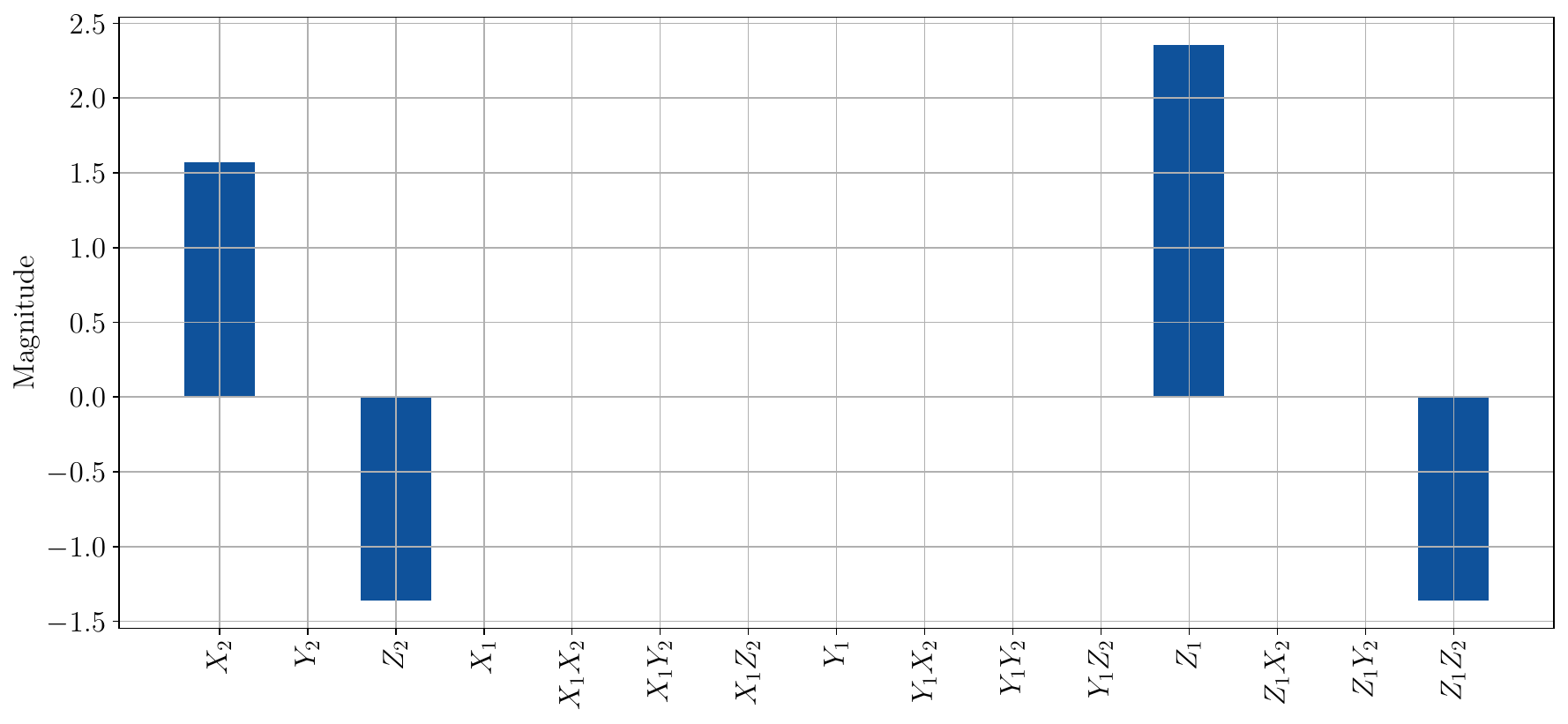}}

    \caption{Bar plot of parameter magnitudes for CNOT gate with (a) 2-local restrictions, and (b) anisotropic Heisenberg restrictions.}
    \label{fig:cnot_minimal_time_params}
\end{figure}

\subsection{Toffoli}
The minimal time Toffoli gates found with 1000 instances of the geodesic algorithm.
\begin{figure}[htb!]
    \centering
    \subfloat[Restriction set $\mathcal{H}_\textrm{2-local}$]{\includegraphics[width=0.85\textwidth]{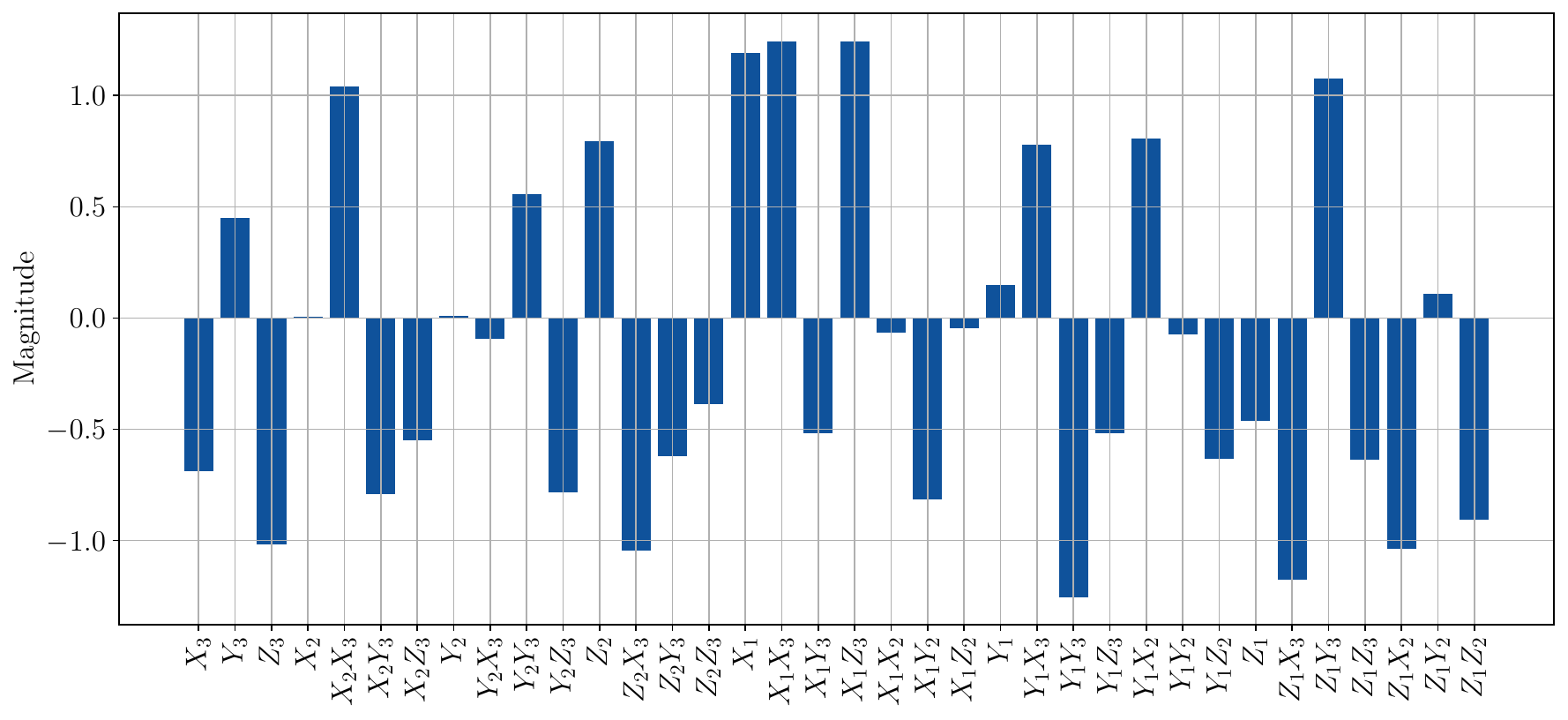}}
    
    \subfloat[Restriction set $\mathcal{H}_\textrm{Heis}$]{\includegraphics[width=0.85\textwidth]{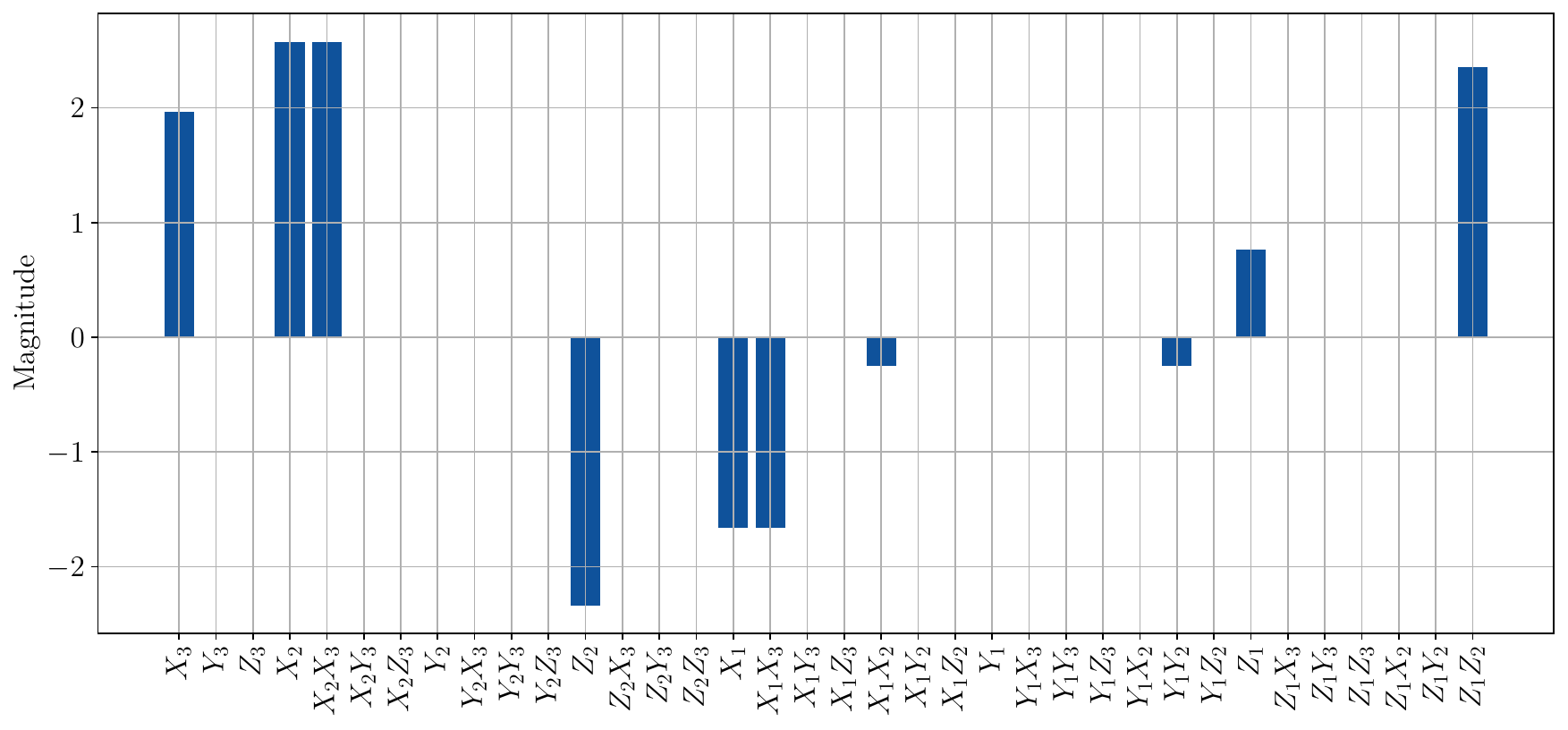}}

    \caption{Bar plot of parameter magnitudes for Toffoli gate with (a) 2-local restrictions, and (b) anisotropic Heisenberg restrictions.}
    \label{fig:toffoli_minimal_time_params}
\end{figure}

\subsection{Fredkin}
The minimal time Fredkin gates found with 1000 instances of the geodesic algorithm.
\begin{figure}[htb!]
    \centering
    \subfloat[Restriction set $\mathcal{H}_\textrm{2-local}$]{\includegraphics[width=0.85\textwidth]{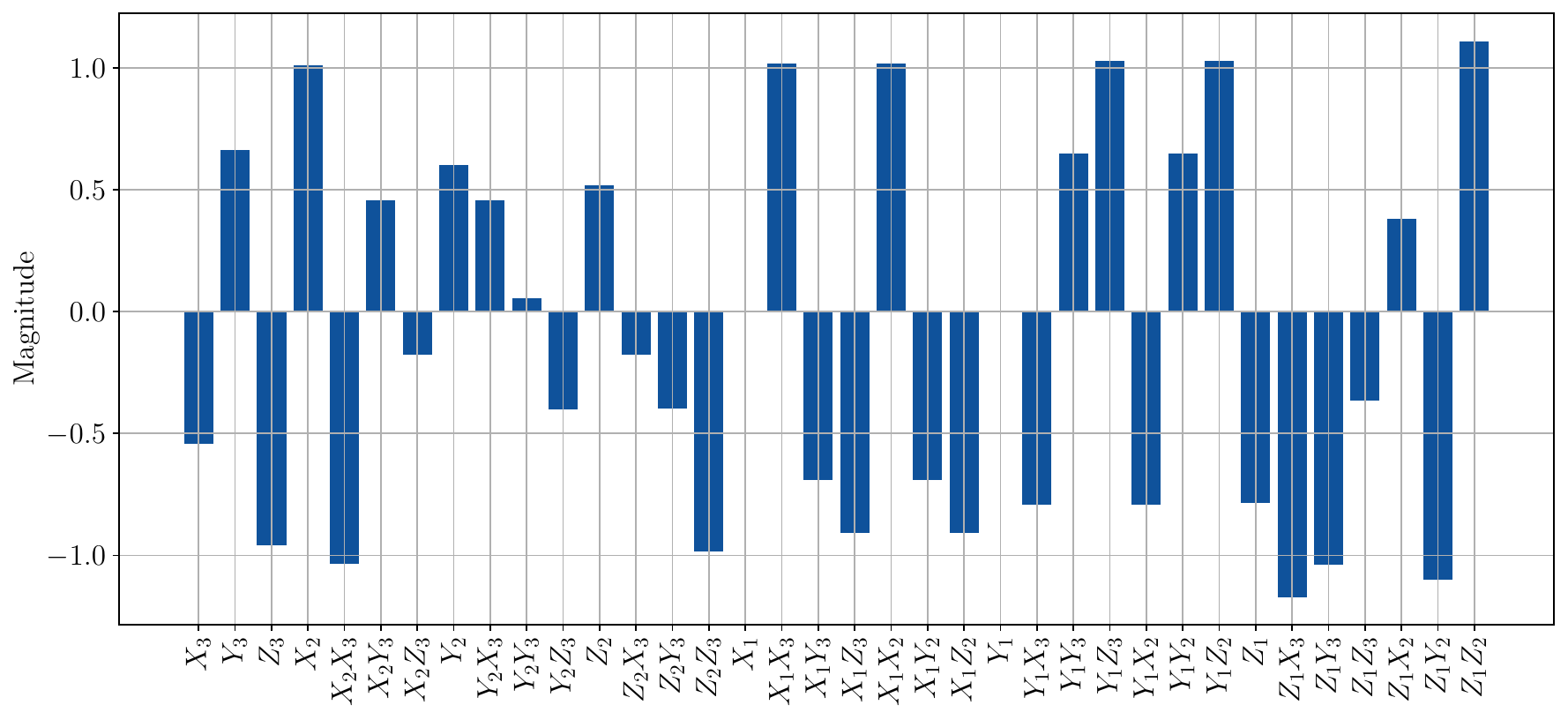}}
    
    \subfloat[Restriction set $\mathcal{H}_\textrm{Heis}$]{\includegraphics[width=0.85\textwidth]{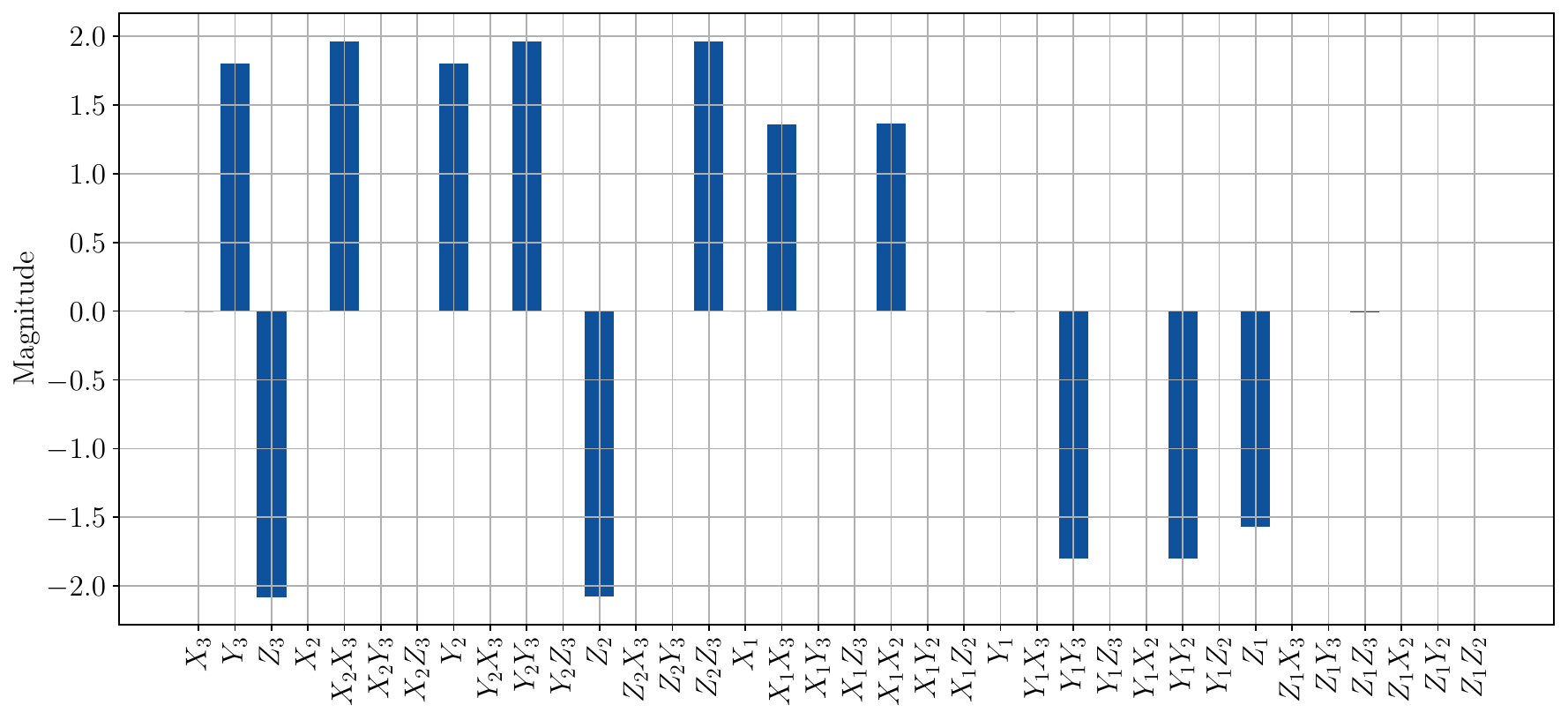}}

    \caption{Bar plot of parameter magnitudes for Fredkin gate with (a) 2-local restrictions, and (b) anisotropic Heisenberg restrictions.}
    \label{fig:fredkin_minimal_time_params}
\end{figure}

\end{document}